\pdfoutput=1
\documentclass[apjl]{emulateapj}

\usepackage{epstopdf}
\usepackage{lineno}
\usepackage{xspace}
\usepackage{graphicx}
\usepackage{amssymb}
\usepackage{amsmath}
\usepackage{natbib}

\shorttitle{Detection of QPOs in OJ~287}
\shortauthors{Bhatta et al.}

\begin{document}

\title{Detection of Possible Quasi-periodic Oscillations \\ in the Long-term Optical Light Curve of the BL Lac Object OJ~287}

\author{
G.~Bhatta\altaffilmark{1},
S.~Zola\altaffilmark{1,\,2},
{\L}.~Stawarz\altaffilmark{1},
M.~Ostrowski\altaffilmark{1},
M.~Winiarski\altaffilmark{2\dag},
W.~Og{\l}oza\altaffilmark{2},
M.~Dr\'o\.zd\.z\altaffilmark{2},
M.~Siwak\altaffilmark{2},
A.~Liakos\altaffilmark{3},
D.~Kozie\l-Wierzbowska\altaffilmark{1},
K.~Gazeas\altaffilmark{4},
B.~Debski\altaffilmark{1},
T.~Kundera\altaffilmark{1},
G.~Stachowski\altaffilmark{2},
and V.~S.~Paliya\altaffilmark{5}
}

\altaffiltext{1}{Astronomical Observatory of the Jagiellonian University, ul. Orla 171, 30-244 Krak\'ow, Poland}
\altaffiltext{2}{Mt. Suhora Observatory, Pedagogical University, ul. Podchorazych 2, 30-084 Krak\'ow, Poland}
\altaffiltext{3}{Institute for Astronomy and Astrophysics, Space Applications and Remote Sensing, National Observatory of Athens, Penteli, Athens, Greece}
\altaffiltext{4} {Department of Astrophysics, Astronomy and Mechanics, National \& Kapodistrian University of Athens, Zografos GR-15784, Athens, Greece}
\altaffiltext{5} {Department of Physics and Astronomy, Clemson University, Kinard Lab of Physics, Clemson, SC 29634- 0978, USA}

\email{email: {\tt gopalbhatta716@gmail.com}}

\begin{abstract}
Detection of periodicity in the broad-band non-thermal emission of blazars has so far been proven to be elusive. However, there are a number of scenarios which could lead to quasi-periodic variations in blazar light curves. For example, orbital or thermal/viscous period of accreting matter around central supermassive black holes could, in principle, be imprinted in the multi-wavelength emission of small-scale blazar jets, carrying as such crucial information about plasma conditions within the jet launching regions. In this paper, we present the results of our time series analysis of $\sim 9.2$ year-long, and exceptionally well-sampled optical light curve of the BL Lac OJ 287. The study primarily uses the data from our own observations performed at the Mt. Suhora and Krak\'ow Observatories in Poland, and at the Athens Observatory  in Greece. Additionally, SMARTS observations were used to fill in some of the gaps in the data. The  Lomb-Scargle Periodogram and the Weighted Wavelet Z-transform methods were employed to search for the possible QPOs in the resulting optical light curve of the source. Both the methods consistently yielded possible quasi-periodic signal around the periods of $\sim 400$ and $\sim 800$ days, the former one with a significance (over the underlying colored noise) of $\geq 99\%$. A number of likely explanations for such are discussed, with a preference given to a modulation of the jet production efficiency by highly magnetized accretion disks. This supports the previous findings and the interpretation reported recently in the literature for OJ 287 and other blazar sources.
\end{abstract}


\section{Introduction \label{sec:intro}}

Quasi-periodic oscillations (QPOs) are routinely found in the X-ray light curves of Galactic binary systems \citep[e.g.,][for a review]{Remillard06}. In radio-quiet AGN, the first detection of short ($\sim 1$\,h) X-ray QPOs was reported by \citet{Gierlinski08} for the narrow-line Seyfert 1 galaxy REJ\,1034+396, followed by subsequent discoveries \citep[e.g.,][]{Alston15}. In the optical domain, 5.2\,yr periodicity has recently been found in the light curve of the radio-quiet quasar PG\,1302-102 \citep{Graham15a}, and some other analogous candidates have been identified (\citealt{Graham15b}; see also \citealt{Liu15} for the case of the high-redshift radio-loud quasar PSO\,J334.2028+01.4075, and \citealt{Zheng16} for the radio-quiet quasar SDSS\,J0159+0105).

Short time scale quasi-periodic oscillations in optical or X-ray blazar light curves --- with characteristic timescales of the orders of hours or even minutes --- have been claimed in the past in several cases (e.g., \citealt{Lachowicz09} for PKS\,2155-304; e.g.  \citealt{Rani10} for S5\,0716+714); these results have not been confirmed for the other periods in the same sources. On the other hand, analogous findings regarding quasi-periodicity in blazar light curves on longer timescales of months or years seem more robust, as summarized below.

At radio frequencies, harmonics with periods ranging from about one year up to several years or a decade have been reported for AO\,0235 (\citealt{Liu06}; see also \citealt{Raiteri01}), PKS\,1510-089 \citep{Xie08}, NRAO\,530 \citep{An13}, and PKS\,1156+295 (\citealt{Wang14}, who confirmed the previous analyses by \citealt{Hovatta07,Hovatta08}). Also, \citet{King13} reported persistent $\sim 150$\,days periodicity in the 15\,GHz light curve of J1359+ 4011. Similar results have been reported in the optical domain, including \citet{Sandrinelli14,Sandrinelli16}, who confirmed the $\sim 315$\,days period in the light curve of PKS\,2155-304 claimed previously by \citet{Zhang14}, or \citet{Sandrinelli15,Sandrinelli16} who found marginally significant ($\sim 3\sigma$) periodicity in PKS\,0537-441, OJ\,287, 3C\,279, PKS\,1510-089 and PKS\,2005-489, on timescales ranging from tens of days up to a few years (often in harmonic relations).

In the X-ray domain, \citet{Rani09} found the $\sim 17$\,days and $\sim 420$\,days quasi-periodicity in AO\,0235+164 and 1ES\,2321+419, respectively. In the high-energy $\gamma$-ray regime, \citet{Sandrinelli14} first detected the $\sim 635$\,day period in the light curve of PKS\,2155-304, most likely in a harmonic relation with the QPO found in the source at optical frequencies. More recently, \citet{Ackermann15} reported a periodic behavior of the blazar PG\,1553+113 with the characteristic timescale of 2.2\,yr (similar to the one found at optical frequencies, but not in the radio domain).

BL Lac object OJ~287 (RA=08h\ 54m\ 48.87s, Dec= +20d\ 06m\ 30.64s and $z = 0.306$) is by now one of the best studied blazars in the history. Sparsely monitored for more than a century, and intensely observed during the last $40-50$ years, it has become the most famous case of a blazar periodicity, with its double optical outbursts repeating every $\sim 12$\,yr \citep{Sillanpaa88,Hudec13}. Besides this, a number of authors have also claimed the presence of QPOs in the source at various timescales: a $\sim 40-50$\,days quasi-periodicity for the source has been claimed at radio frequencies by \citet{Wu06}, and at optical frequencies by \citet{Pihajoki13}. Also \citet{Sandrinelli16} reported to have found a marginally-significant ($\sim 3\sigma$) signal in the optical light curve of OJ\,287 with the period of $\sim 435$\,days, in addition to a much less significant one of 203\,days.

In this paper, we present our analysis and results of a search for quasi-periodical oscillations in the $9.2$\,year-long and exceptionally densely-sampled optical light curve of the blazar OJ 287, utilizing several widely-used statistical methods, including the Lomb-Scargle Periodogram, and Weighted Wavelet Z-transform. The results of the analysis are given below.

\section{Observations and Data Reduction \label{sec:obs}}

We started optical monitoring of the blazar OJ~287 in September 2006 at three sites: Mt. Suhora Observatory of the Pedagogical University in Krakow (SUH), the Astronomical Observatory of the Jagiellonian University (KRK) and the Astronomical Observatory of the University of Athens (GR). 

Here we report the data taken until the end of the 2014/2015 observing season. Most of the observations were gathered by taking a series of about 10 frames each night from which nightly averages were calculated. We obtained 562 points at SUH, 164 at GR and 123 at KRK. All three sites are equipped with CCD cameras and sets of wide band filters \citep{Bes90}. We took most of the data in the $R$ filter, occasionally also measuring the target colours at SUH and KRK. After calibrating images for bias, dark and flat-field (usually taken on the dusk or dawn sky) with IRAF, we extracted magnitudes with the C-Munipack \citep{hroch98,motl04} software. All scientific images have been visually inspected. We used GSC~1400-222 ($R=13.74^m$) as the comparison star, for which the constancy was checked against the nearby star GSC~1400-444. 

To fill some of the gaps in our observations, optical observations in the $R$ filter from the Small and Moderate Aperture Research Telescope System (SMARTS) were also included; the details of data acquisition and reduction procedures are discussed in \citet{Bonning12}. To match our observations with those from SMARTS, a small offset of 0.02 magnitude, comparable to instrumental errors, was subtracted from the SMARTS data. Further, the  magnitudes were converted into flux units by using the zero points for the filter R given in \citet{Bessell98}.

\begin{figure}[t!]
\begin{center}
\includegraphics[width=\columnwidth,angle=0]{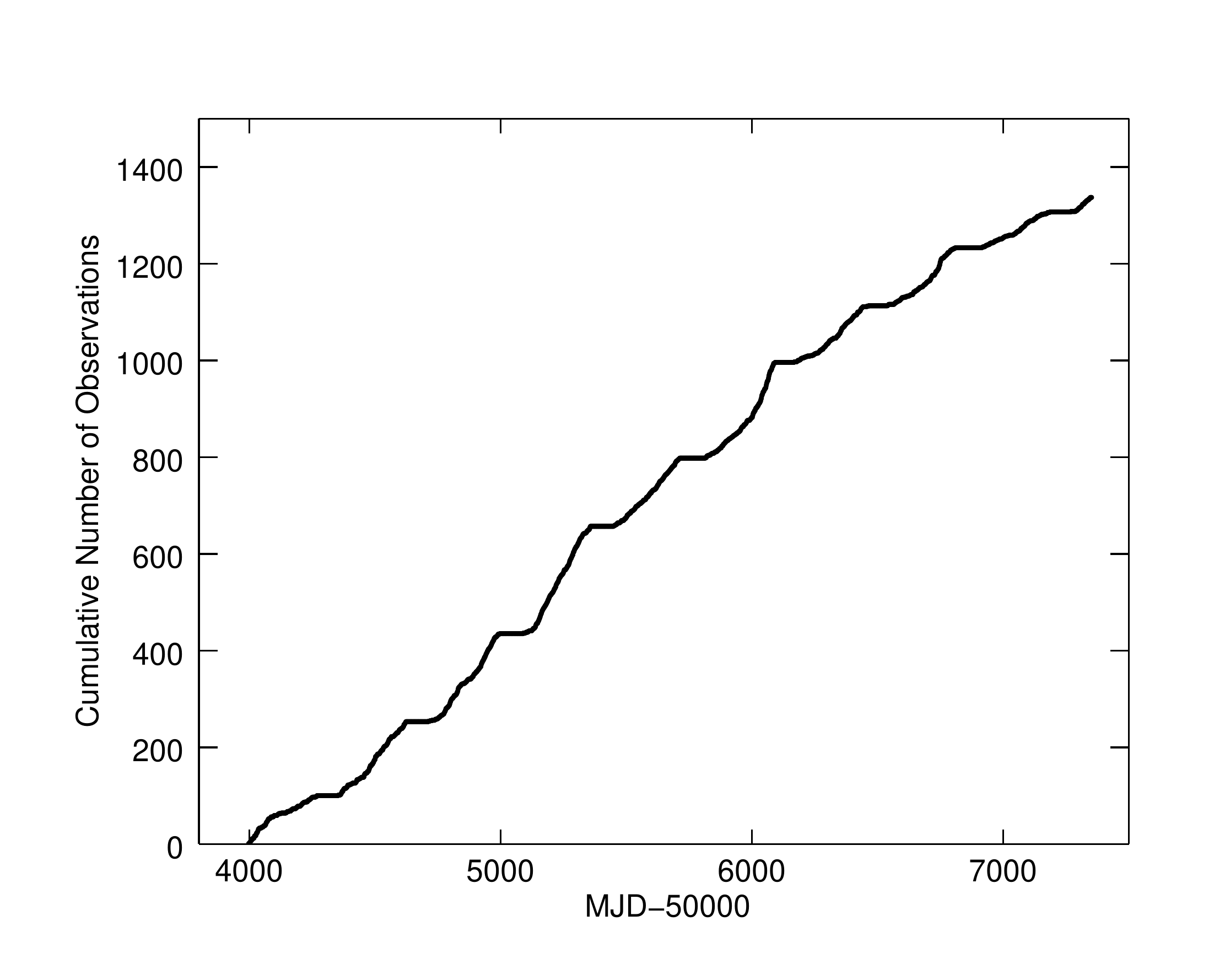}
\caption{The cumulative number of observations in the $R$ band light curve of the blazar OJ~287 as a function of time.}
\label{cum_number}
\end{center}
\end{figure}

\section{Analysis and Results \label{sec:dis}}

\subsection{The Light Curve}

The analyzed optical light curve of the source spans about 9.2\,years, from 2006 till 2015 (53997.590--57352.578~MJD), with a total of 1338 high quality observations. The mean brightness and the standard deviation during the observation period are 14.47 and 0.48 magnitudes, respectively. Excluding 9 summer gaps (for a total of 799 days) due to the source visibility, the average data sampling rate is $\sim$1.92/day. The data acquisition rate indicated by the cumulative number of observations along with the location of the summer gaps are shown in Figure~\ref{cum_number}. The observed nearly-linear increase in the cumulative number of observations with time, except for a few shorter time intervals manifesting in the graph as flat plateaus, is clearly an indicator of the highly regularly-sampled data collected for the whole observation epoch. 

\begin{figure*}[t!]
\begin{center}
\includegraphics[width=\textwidth,angle=0]{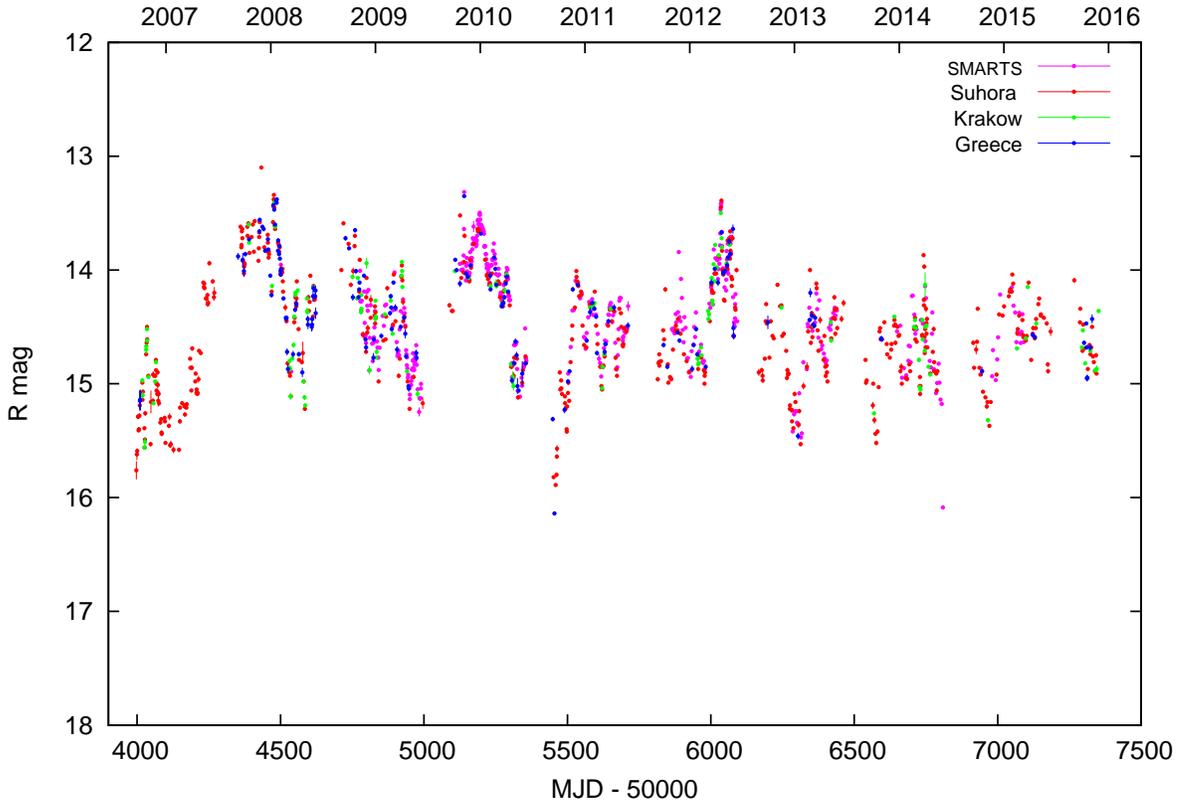}
\caption{The long-term optical ($R$ band) light curve of the blazar OJ~287 from various observatories as indicated on the plot. }
\label{fig:lightcurve}
\end{center}
\end{figure*}

As revealed by the resulting source light curve presented in Figure~\ref{fig:lightcurve}, the blazar exhibits considerable variability during the period: the amplitude of peak-to-peak oscillations \citep{Heidt96} is calculated to be 3.00\,mag, while the fractional variability \citep{vau03,Edelson02} is $44.25 \pm 0.04$\%.

\subsection{Periodicity Search}
\subsubsection{Lomb-Scargle Periodogram}

\begin{figure*}[t!]
\begin{center}
\begin{tabular}{cc}
\includegraphics[angle=0,width =\columnwidth]{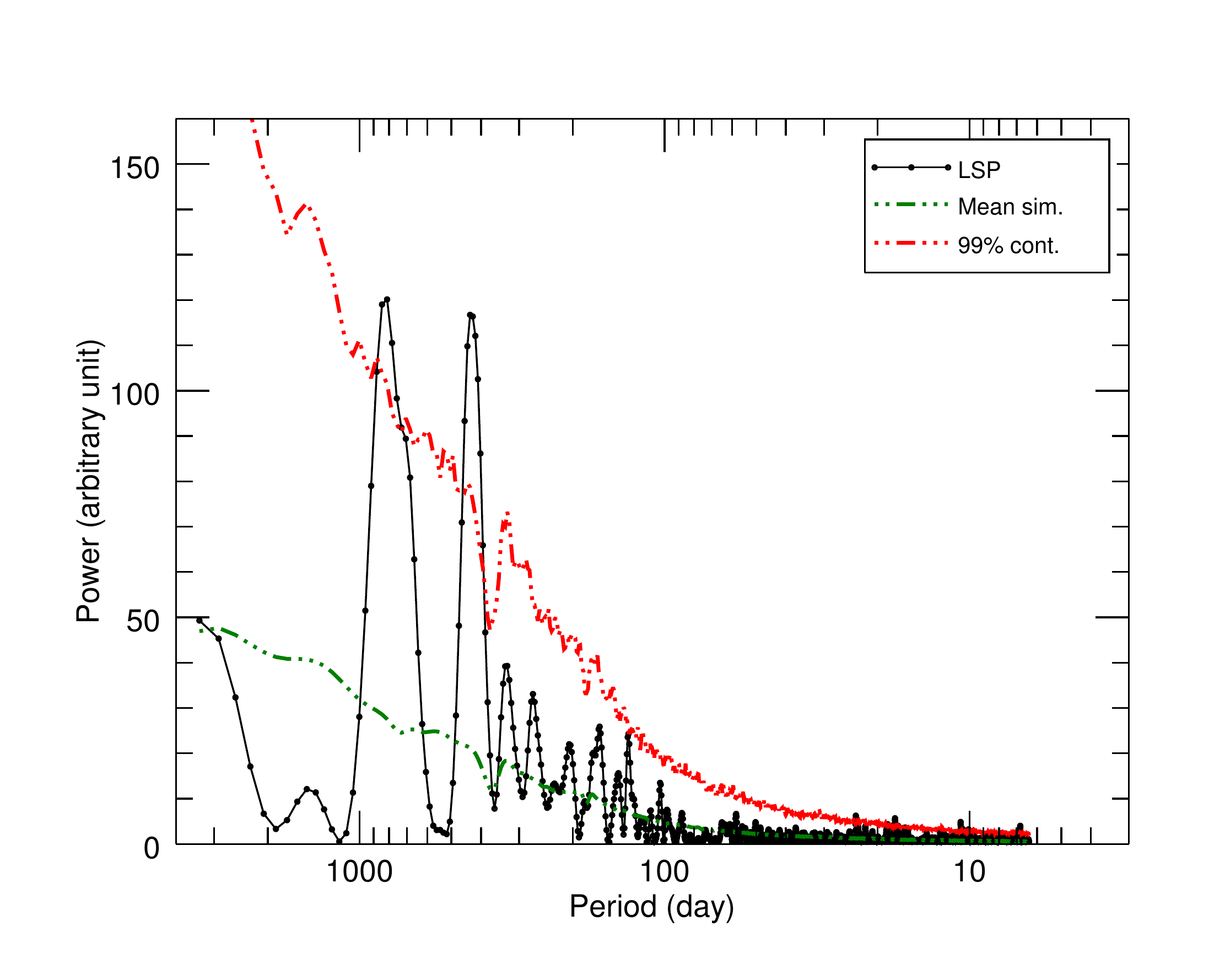}&
\includegraphics[angle=0,width =\columnwidth]{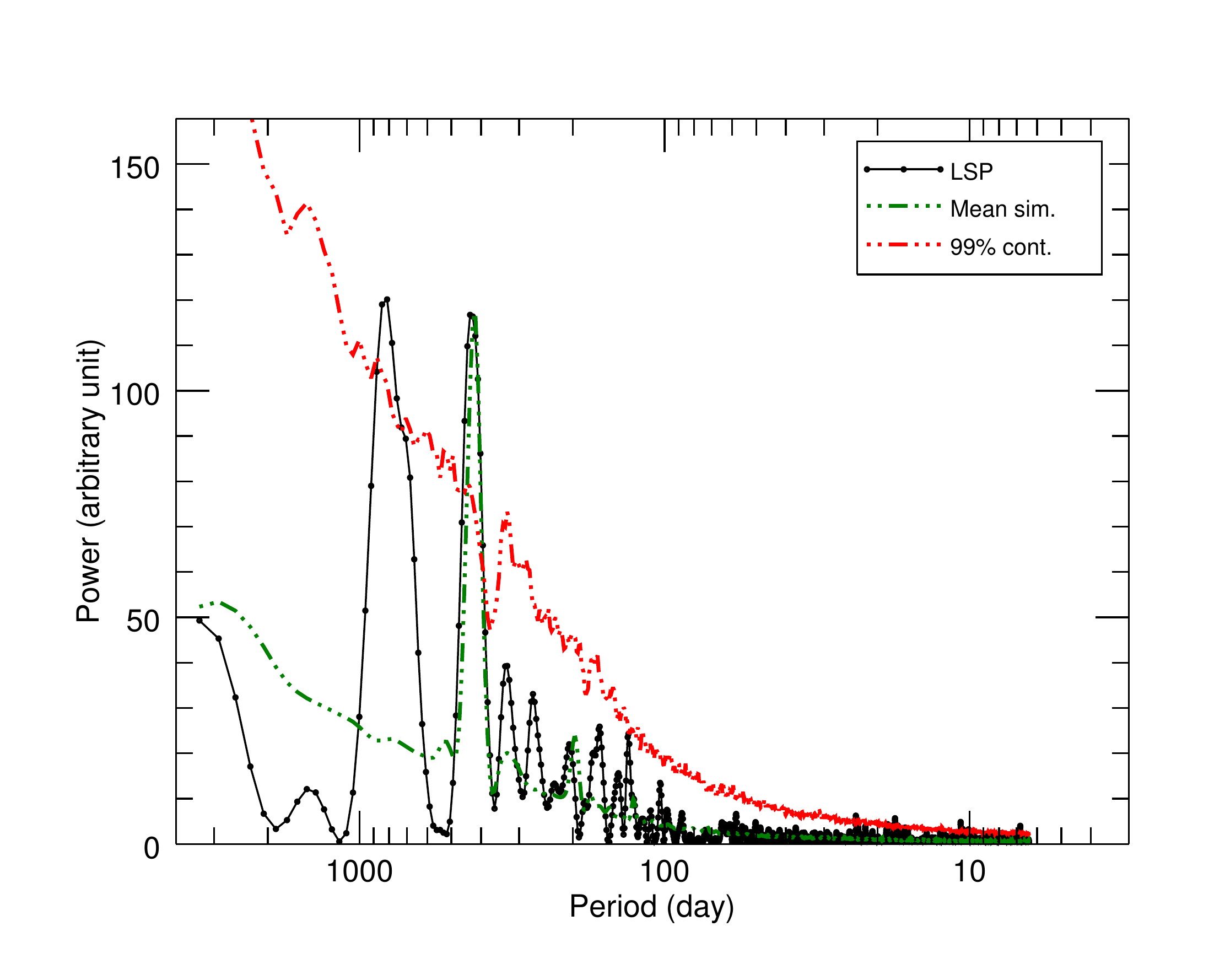}\\
\includegraphics[angle=0,width =\columnwidth]{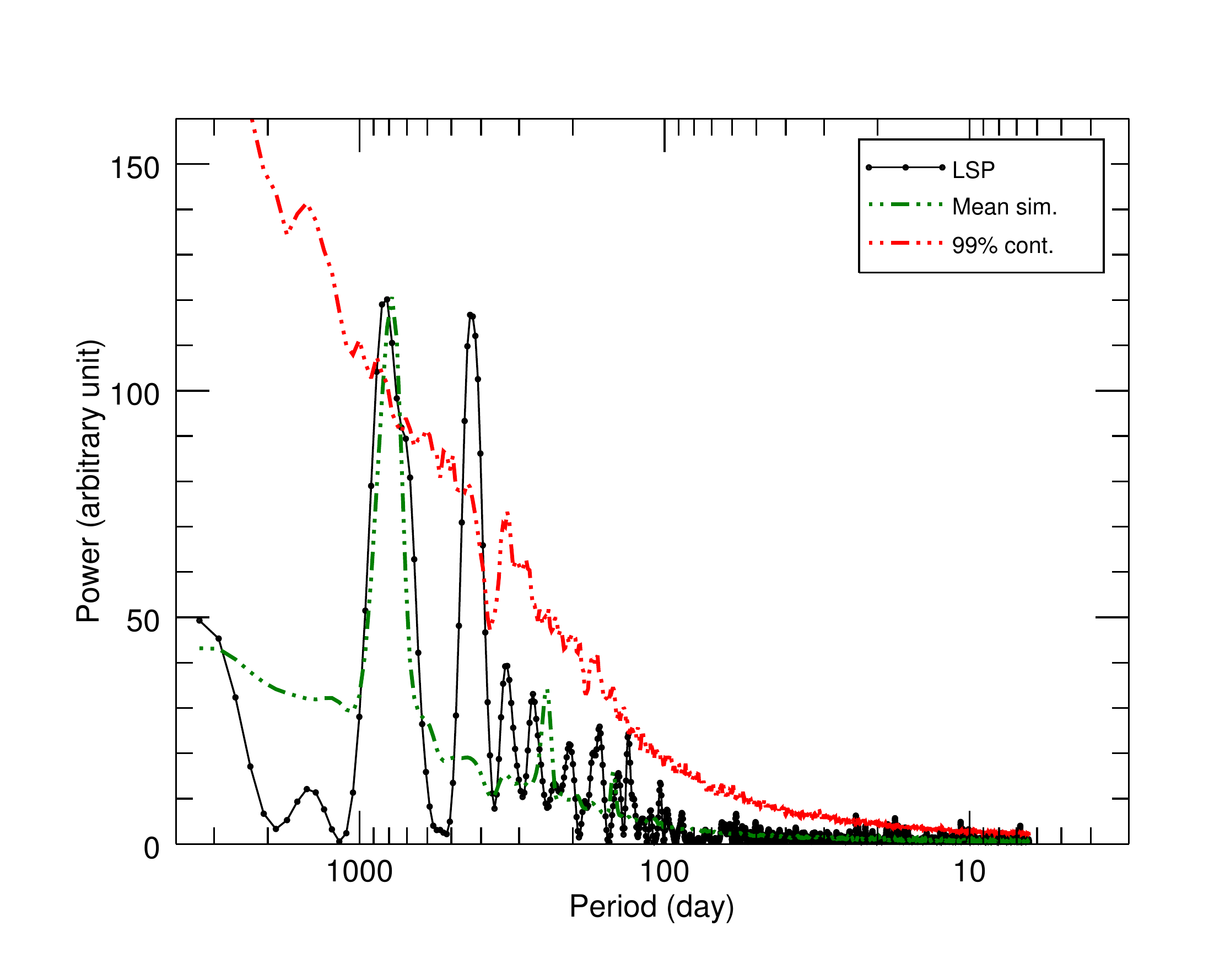}&
\includegraphics[angle=0,width =\columnwidth]{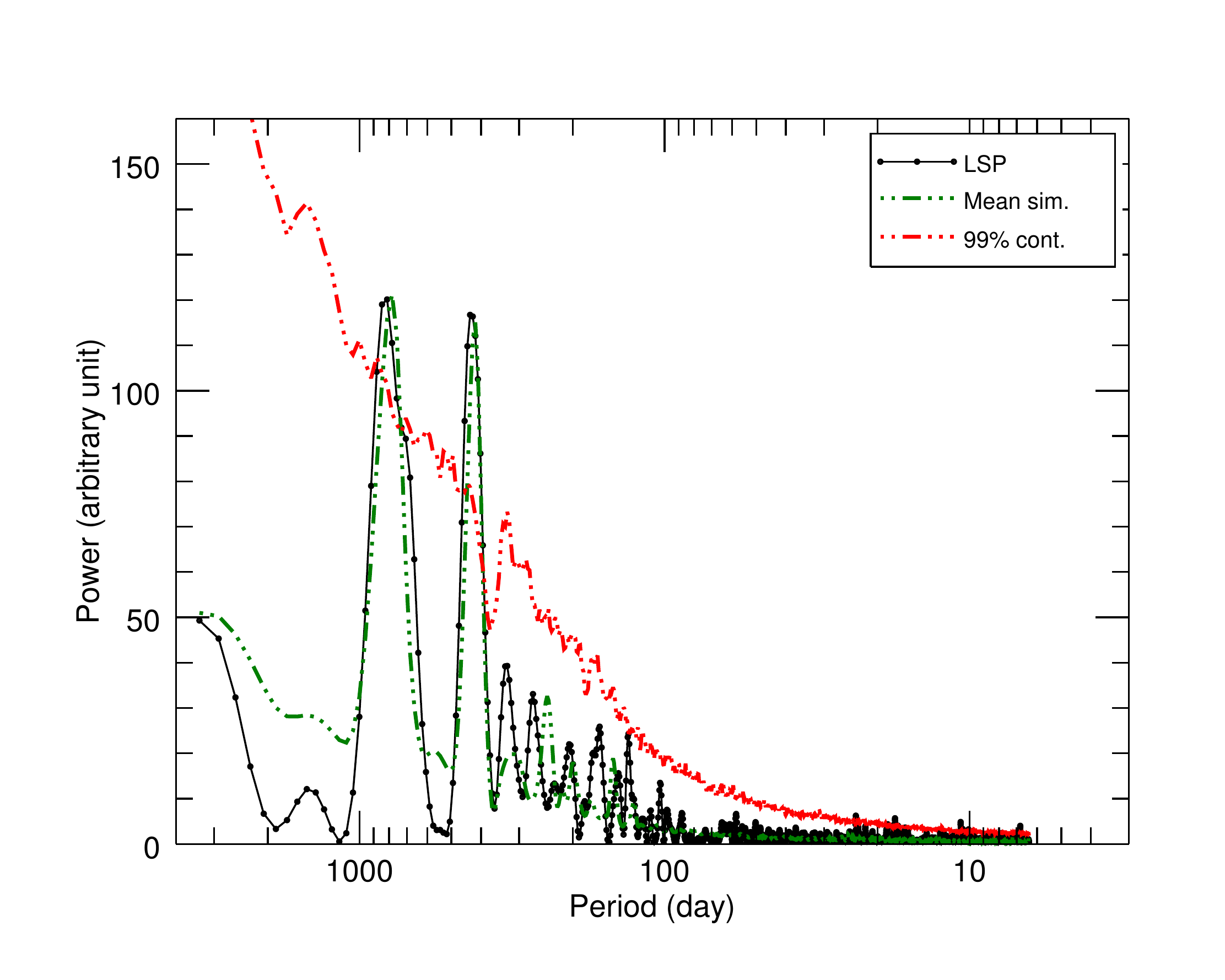}\\
\end{tabular}
\caption{The LSP of the optical $R$ band flux of OJ~287 (black), along with that of the mean of $1,000$ simulated light curves ( green) and the corresponding 99\% confidence contour (red). The simulated periodograms involve a  broken power-law (top-left panel), a  broken power-law plus a sinusoidal wave with a period of $410\,$days (top-right), a  broken power-law plus a sinusoidal wave with a period of $789$ days (bottom-left), and finally a broken power-law plus sinusoidal waves with periods of $410$ and $789$ days (bottom-right).}
\label{LSP}
\end{center}
\end{figure*}

The Lomb-Scargle Periodogram (LSP), one of the most widely used methods in the QPO analysis, modifies the periodogram such that the least-square fitting of sine waves of the form $X_{f}(t)= A \cos\omega t +B \sin\omega t$ is minimized \citep{Lomb76,Scargle82}. The \textit{modified periodogram} is expressed as
\begin{equation}
P=\frac{1}{2} \left\{ \frac{\left[ \sum_{i}x_{i} \cos\omega \left( t_{i}-\tau \right) \right]^{2}}{\sum_{i} \cos^{2}\omega \left (t_{i}-\tau \right) } + \frac{\left[ \sum_{i}x_{i} \sin\omega \left( t_{i}-\tau \right) \right]^{2}}{\sum_{i} \sin^{2}\omega \left( t_{i}-\tau \right)} \right\} \, ,
\label{modified}
\end{equation}
where $\tau$ is given by
\begin{equation}
\tan\left( 2\omega \tau \right )=\frac{\sum_{i} \sin\omega t_{i}}{\sum_{i} \cos\omega t_{i}} \, .
\end{equation}
This weights the data-points and hence accounts for the problem with unevenly spaced dataset. The calculated LSP of the source light curve in the $R$ band is presented in Figure~\ref{LSP}, which clearly shows two prominent features around the periods of $410\pm38$ and $789\pm58$ days. The associated uncertainties are the half-widths at half-maxima (HWHM) of Gaussian fits to the corresponding periodogram peaks after subtracting the mean periodogram due the colored noise determined from the Monte Carlo simulations, as explained in detail in section \ref{Significance}.

\subsubsection{Weighted Wavelet Z-transform}

Although the LSP method accounts for irregular spacing in a time series, the method does not take into account time fluctuations in the periodic signal information. In real astronomical systems, quasi-periodic oscillations may develop and evolve in frequency and amplitude over time. In such cases, the wavelet transform method proves to be a more useful tool, and indeed it is frequently employed in the analysis of blazar sources \citep[e.g.,][]{bhatta13,Hovatta08}. 

The method attempts to fit sinusoidal waves localized in both time and frequency domains, which can be scaled in frequency and shifted in time \citep{Torrence98}. However, for unevenly spaced data the method still may not yield reliable results due to the fluctuations of the local number density of the data-points, an important parameter in the transform. As a result, a false time evolution of the characteristic frequency may be recovered. In this context, \citet{Foster96} suggested that the problem can be handled efficiently by rescaling of the wavelet functions such that they can be viewed as weighted projections on the trial functions $\phi_{1}\!\left( t \right) =\mathbf{1}\!\left( t \right)$, $\phi _{2}=\cos\left[ \omega (t-\tau ) \right]$, and $\phi _{2}=\sin\left[ \omega (t-\tau ) \right]$ by the weight given as $w_{i}=e^{-c \, \omega^{2}\left( t_{i}-\tau \right)^{2}}$, where $c$ acts as a fine tuning parameter usually chosen around $0.0125$. This modified approach is known as the weighted wavelet Z-transform (WWZ). With $V_{x}$ and $V_{y}$ as weighted variations of the data and the model function, respectively, the WWZ power can be written as,
\begin{equation}
\label{wwz}
WWZ=\frac{\left ( N_{eff}-3 \right )V_{y}}{2\left ( V_{x}-V_{y} \right )},
\end{equation}
where $N_{eff}$ is the effective number of the data points \citep[for further details see][]{Foster96}.

\begin{figure}[t!]
\begin{center}
\includegraphics[width=\columnwidth,angle=0]{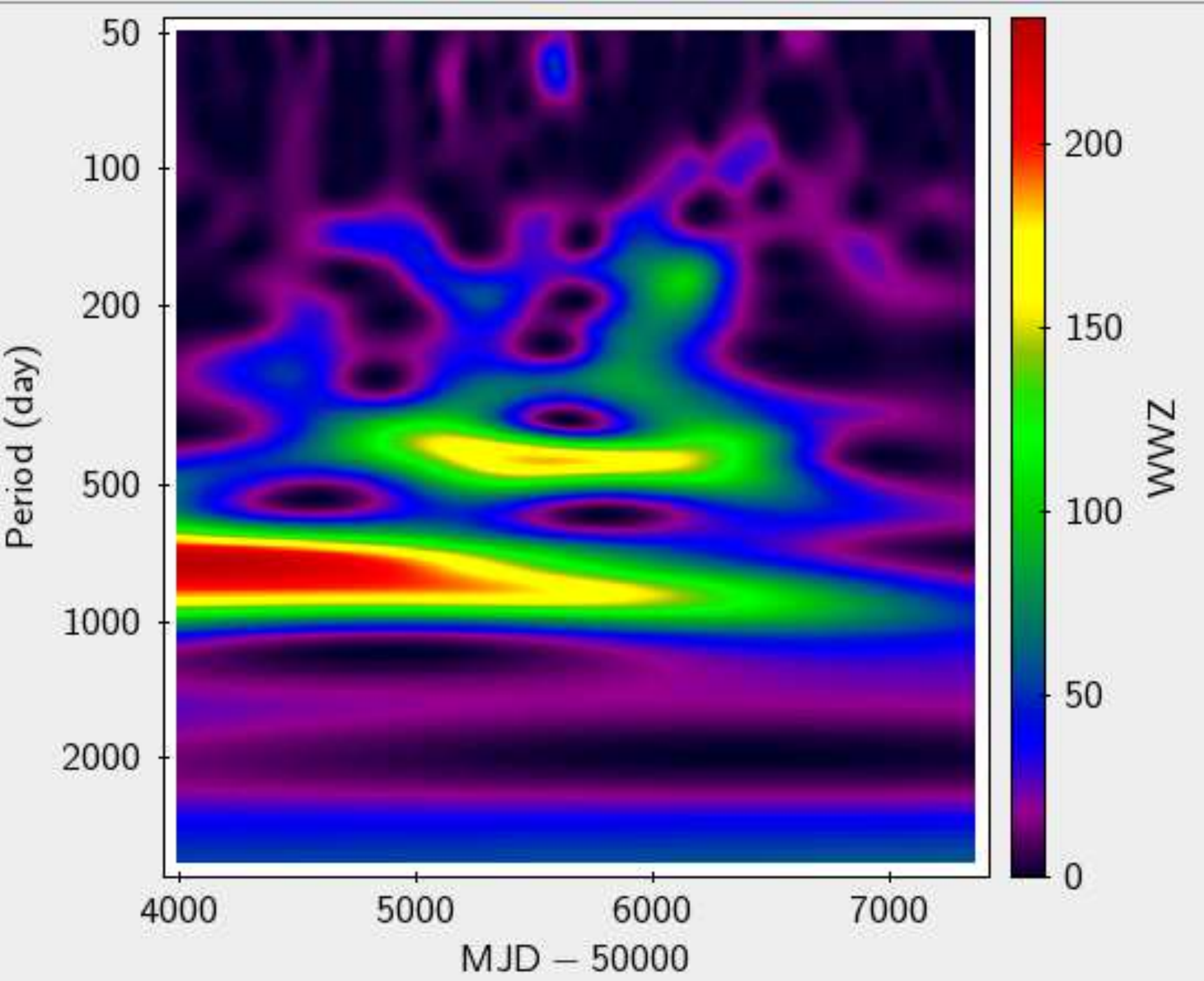}\\
\includegraphics[width=\columnwidth,angle=0]{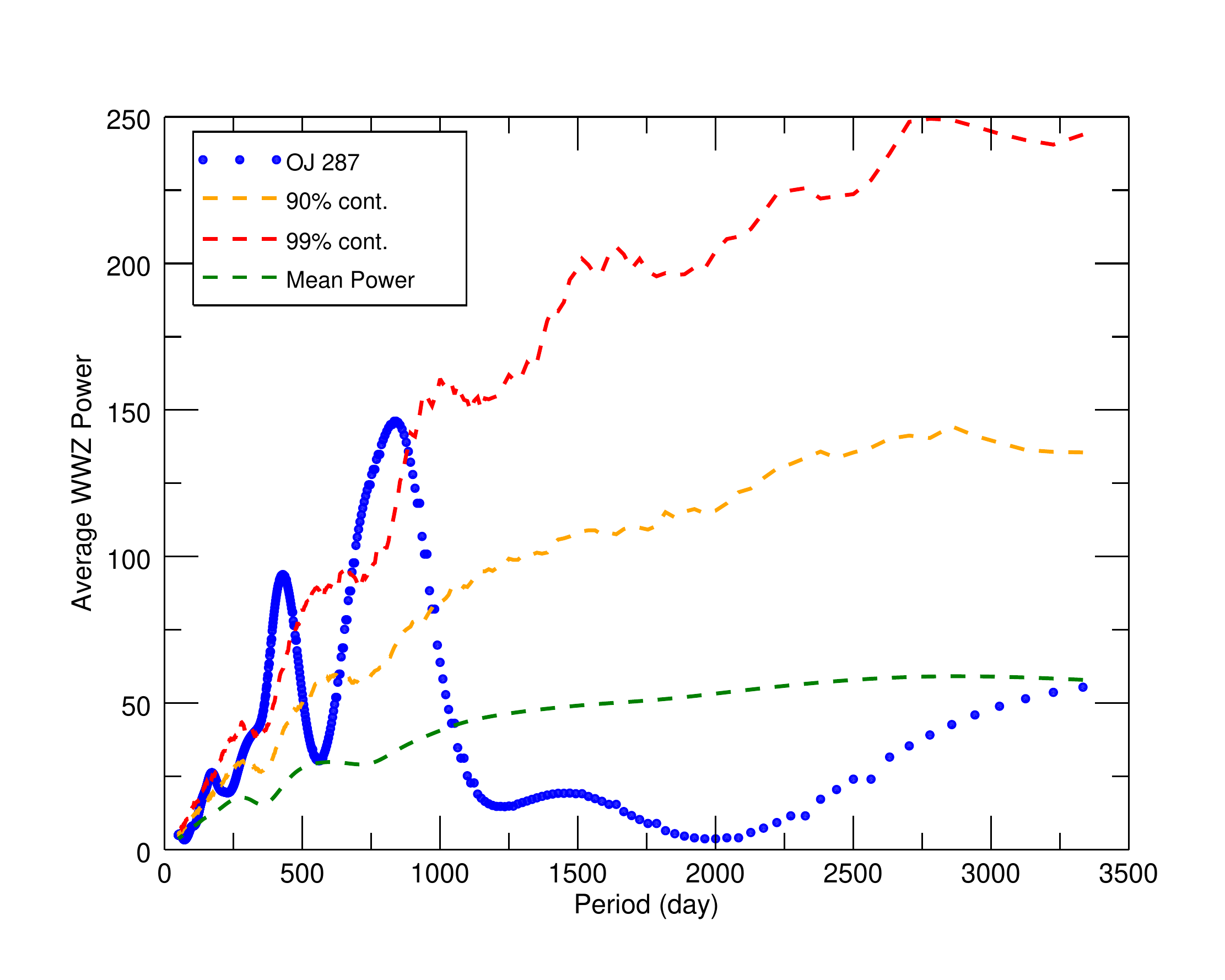}
\caption{{\it Top panel}: The color-scaled WWZ power of OJ~287 in the time-period plane. {\it Bottom panel}:  Time ($\tau$)-averaged WWZ power as a function of period (blue symbols), mean time ($\tau$)-averaged WWZ power (green dashed-line), and $90\%$ and $99\%$ confidence contour (orange and red dashed-lines, respectively) from the simulations.}
\label{wavelet}
\end{center}
\end{figure}

We computed the WWZ power of the $R$ band flux of OJ~287 as a function of time and characteristic period, with the conventional value $c=0.0125$, by using the WWZ software available at American Association of Variable Stars Observers (AAVSO)\footnote{\texttt{https://www.aavso.org/software-directory}}.  The top panel in Figure~\ref{wavelet} presents the color-scaled WWZ power of the source in the time-period plane. The figure indicates that the power for the characteristic periods centered at $419\pm52$ days and $817\pm130$ days is highly significant, confirming the result obtained using the LSP. The uncertainties given here for the periods are evaluated as the means of the HWHMs of the Gaussian fits centered around a maximum power along the time ($\tau$) axis; these uncertainties are relatively large because the maximum powers ``meander'' along $\tau$. 

The QPO corresponding to the period of $817\pm130$ days starts from the beginning of the observing epoch and gradually dies out towards the end of the observations, although we note that the high power at the beginning could have been affected to some extent by the edge effects. On the other hand, a QPO behavior around the period of $419\pm52$ days shows up a little further from the start of the campaign, and decays well before the end of observations. The bottom panel of Figure~\ref{wavelet} presents the average oscillation power at a given period. As shown, while in the lower frequency domain the signal seemed weaker and transitory, at higher frequencies the signal becomes more coherent and persistent.

\section{Significance and Uncertainty Estimation}
\label{Significance}

Although the above described LSP and WWZ methods are widely used in the time series analysis in astrophysics, the effects related to uneven sampling of a light curve --- which can sometimes produce spurious peaks in the periodogram that can be mistaken for a real signal --- should be analyzed in detail, especially when searching for QPO oscillations. In addition, there is a pitfall that blazar variability is, in general, of a colored noise type, with larger amplitude fluctuations at longer variability timescales; with such, high amplitude peaks in the lower frequency region of the periodograms can sometimes pose as QPO features \citep[see in this context][]{Press78}. Hence, in order to establish a robust significance of the potential QPO oscillations revealed by both the LSP and WWZ methods in OJ~287, \emph{both} the uneven sampling of the source light curve \emph{and} the colored noise-type behaviour of the blazar, have to be addressed carefully. This can be done by means of Monte Carlo (MC) simulations of a large number of red-noise light curves of the source, obtained by randomizing both the amplitude and the phase of the Fourier components, following \citet{TK95}. 

First, in order to characterize the exact form of the underlying colored noise of the source (needed as an input for simulating the source light curves), we followed the power response method \citep[PSRESP; see][]{Uttley02}, which is widely used in modeling the AGN power spectra \citep[e.g.,][]{Chatterjee08,Edelson2014,Kapanadze2016,Stone16,bhatta16}. In short, the method attempts to fit the binned periodogram with a model PSD which maximizes the probability that the given model PSD can be accepted.

\begin{figure*}[t!]
\begin{center}
\begin{tabular}{cc}
\resizebox{\columnwidth}{!}{\includegraphics[angle=0]{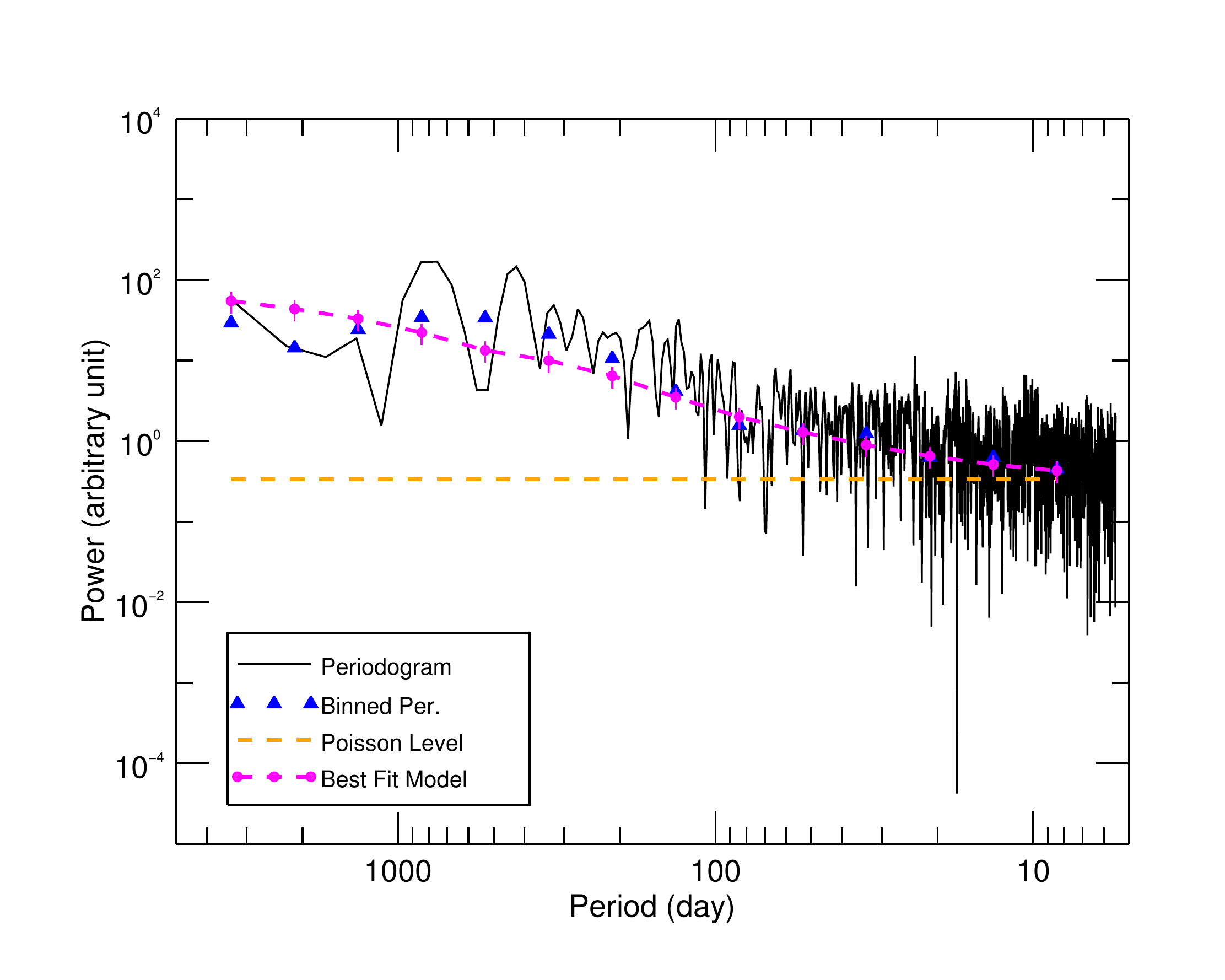}} &
\resizebox{\columnwidth}{!}{\includegraphics[angle=0]{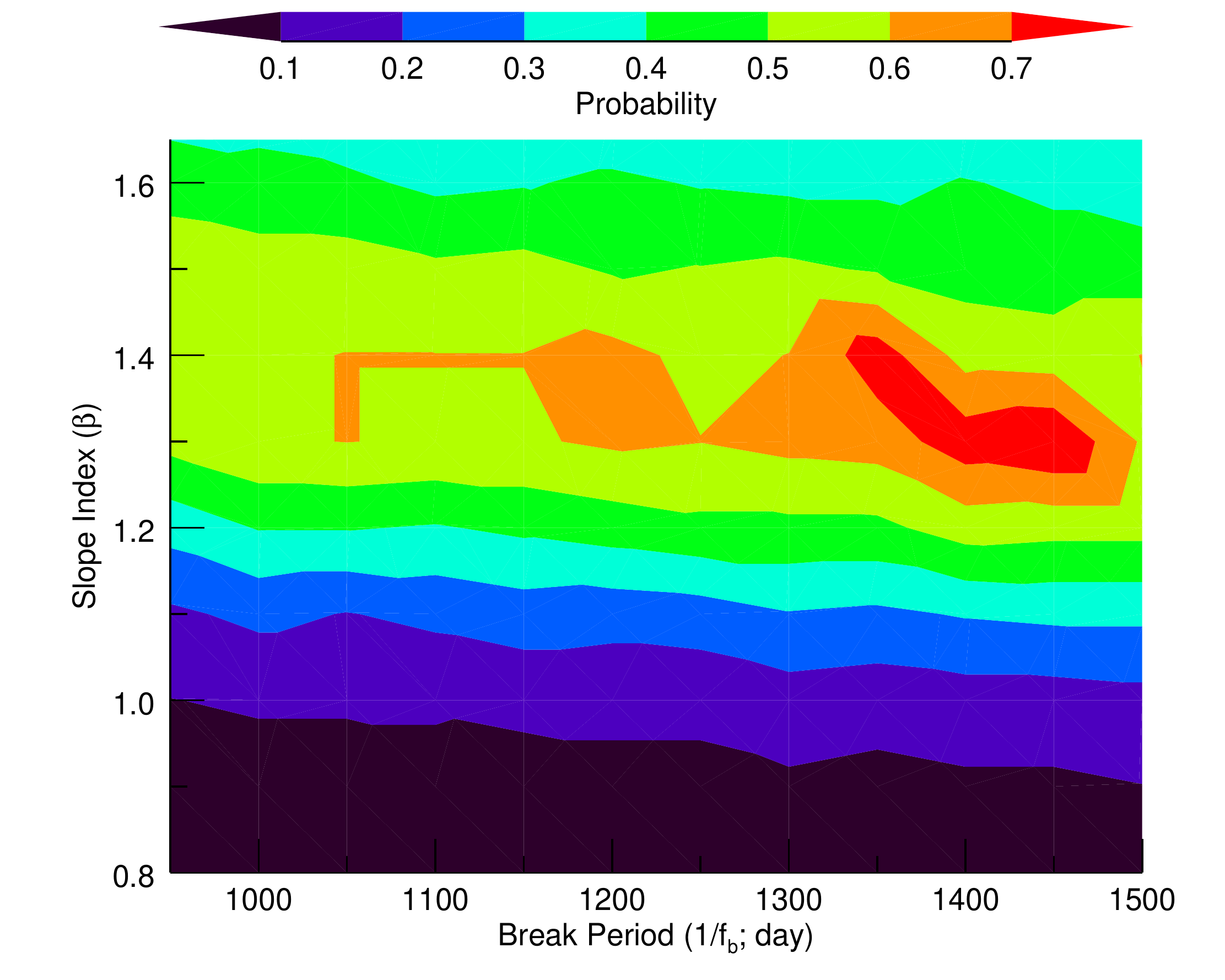}} \\
\end{tabular}
\caption{{\it Left panel}: Unbinned (black curve) and binned (blues triangles) DFT periodogram for the OJ 287 light curve analyzed in this paper. The binned periodogram corresponding to the reference model with spectral slope $\beta =1.3$  and break frequency $f_{b} =1/1400\ d^{-1}$ is shown in magenta and the associated errors are $1 \sigma$ deviations of the distribution in the simulated periodograms at a given frequency. The horizontal orange dashed-line represents the mean Poisson noise level. {\it Right panel}: Color scaled probability contours for various values of the slopes and the break frequencies for model PSD used to simulate the light curves. }
\label{fig:5}
\end{center}
\end{figure*}

The discrete Fourier periodogram (DFP) of the light curve  and the periodogran binned in logarithm frequency  to reduce the scatter, are both presented in the left panel of Figure~\ref{fig:5}; the horizontal orange dashed line marks the Poisson noise level. We attempted to model the periodogram with a broken power-law of the form $P(f) \propto [ 1+(f/f_{b})^{2}]^{-\beta /2}$, where $\beta$ and $f_{b}$ correspond to the spectral slope and the break frequency, respectively. This particular ``knee'' model is designed to suppress the excess power at long variability timescales, given the fact that the power in the lower frequency domain of the periodogram cannot increase indefinitely, in accord with the analysis of the long-term optical light curve of PKS\,2155+304 presented by \citet{Kastendieck2011}, who found the characteristic break timescale in the blazar of the order of a few years.

In order to determine the best fit PSD model, the observed binned discrete Fourier periodogram (blue triangles in the left panel of Figure\,\ref{fig:5}) was compared with the mean periodogram of 1,000 simulated light curves for each pair of trial values of $\beta$ and $f_{b}$. In particular, as a measure of a goodness of the fit, we calculated the probabilities of a model being accepted for 10 trial $\beta$s ranging from 0.8 to 1.7 with the interval of 0.2, and 7 trial $f_{b}$s ranging from 900 to 1500 days with the interval of 100 days (a total of $10 \times 7 \times 1000 = 70,000$ simulated light curves). The resulting distribution of the probabilities for the corresponding trial break periods and PSD slopes are presented in the the color-coded contour plot in the right panel of Figure~\ref{fig:5}. The resulting best-fit model (the highest probability of 0.73), corresponding to $1/f_{b}=1400\pm55$\,days and $\beta = 1.3\pm0.2$, is represented by the magenta points connected by the dashed line in the left panel of Figure~\ref{fig:5}. The uncertainties associated with the slope and the break period were calculated here by taking the half the vertical and horizontal ranges of the highest probability ($>0.7$) region denoted in the right panel of Figure~\ref{fig:5} in red.

 The best-fit PSD model described above was next used to simulate numerous light curves of the source, which were re-sampled according to the observed light curve. The distribution of such 1,000 light curves was studied to establish the significance of the peaks seen in the observed LSP; in particular the 99\% significance contour was determined by the 99th percentile of a distribution of the power in the periodogram of the simulated light-curves. As given in Figure\,\ref{LSP} (top-left panel), this significance turns out to be 99.8\% and 99.3\% for the peak around $410$ and $789$ days, respectively. In order to validate further the QPO detection, sinusoidal oscillations with $419$\,days and $780$\,day periods and a small amplitude ($\sim 0.7 \, \sigma_{flux}$) were superimposed on the previously generated light curves. The corresponding average periodograms resulting from $1,000$ simulations are given by green curves in Figure~\ref{LSP}: 410\,-day period only in the top-right panel, 780\,-day period only in the bottom-left panel, and both periods together in the bottom-right panel.

Similarly, in order to establish the significance of the observed WWZ power of the light curve, the WWZ analysis were carried out on the 1,000 simulated light curves from the above described reference PSD model, and their distribution of  WWZ power averaged over $\tau$ (time; day) in the considered frequency-space (alternatively period-space) were studied for the significance estimates. In particular, the observed average WWZ power was compared against the $90\%$ and $99\%$ confidence contour ($90$th and $99$th percentile at a given frequency) from the distribution of the $\tau$-averaged WWZ power for the simulated light curves represented in the lower panel of the Figure \ref{wavelet} by the orange and red dashed-curves, respectively. The significance of the of the peaks centered around $\sim400$ and $\sim800$ were in this way evaluated as 99.7\% and 99.0\%, respectively. 

\section{Discussion and Conclusion \label{discussion}}

Periodicity in the light curves of blazar sources (and, more generally, of other types of AGN), may be related to the period of a perturbing object in a binary black hole system \citep[eg][]{Lehto96,Graham15a}, or a jet precession caused by either closely orbiting binary black holes or warped accretion disks \citep[see the discussion in, e.g.,][]{Graham15b,Sandrinelli16}. Moreover, coherent helical/non-ballistic motions of relativistic blobs within blazar jets could be responsible for quasi-periodicity of blazar sources \citep[e.g.,][]{Camenzind92, Mohan15}. Finally, jet modulation by various instabilities developing within the innermost parts of accretion disks, in principle, could result in quasi-periodic modulation of the observed jet emission \citep[see the discussion in][]{Liu06,Wang14}.

In the specific context of OJ\,287, previous claims of the detection of (quasi-)periodic oscillations on year-like timescales --- besides the 60-year modulation advocated by \citet{Valtonen06} and the 12-year cycle discussed by \citet{Sillanpaa88} --- have been made by a number authors. For example, \citet{Hughes98} identified a persistent oscillation in the radio light curve of the source, with a characteristic timescale of about $410$ days. This result was supported by the wavelet analysis attempted by \citet{Hovatta08}. More recently, using the data for the partly overlapping observing epoch considered in this paper, \citet{Sandrinelli16} claimed to have detected $\simeq 435\,$d and $\simeq 412\,$d period in the NIR-optical and $\gamma$-ray light curve of the blazar, respectively. The detection of the same signal in more than one band ---  radio, optical, \emph{and} $\gamma$-rays --- enforce the physical relevance of the findings. It is therefore interesting to indicate at this point a possible analogy between a year-like periodicity we see in OJ~287 and the low frequency `C-type' QPOs commonly found in Galactic X-ray binaries. These QPOs come in harmonic peaks sticking out of a power spectrum at frequencies just above/around the transition from the ``white noise'' to the ``pink/red noise'' segments of the PSDs, and are typically linked to the Lense-Thirring precession of the innermost parts of accretion disks \citep[see, e.g.,][]{Stella98,Motta11}.

Considering a possible explanation for the observed QPOs in OJ~287, we note that since binary black holes have been claimed to shape the 12\,yr periodicity of the optical light curve in the source (\citealt{Valtonen08}; but see also \citealt{Villforth10} for a critical review), the same scenario could not account for the $\sim 400$\,d period found in our analysis. Jet precession due to warped accretion disk, on the other hand, as well as a ``grand-design'' helical magnetic field in the OJ\,287 jet, seem both in conflict with the erratic jet wobbling observed on parsec scales by \citet{Agudo12} in the high-frequency radio domain at the time of our monitoring program. Hence we conclude that the likely explanation for the $\sim 400$\,d period in the blazar (with the possibly accompanying $\sim 800$\,d harmonic) could be a jet modulation by the innermost parts of the accretion disk. Interestingly, in the case of magnetically-arrested disks, the characteristic timescale of quasi-periodic oscillations in jet the production efficiency set by the rotating and unstable (``chocking'') magnetic field accumulated at the saturation level around the horizon of a spinning black holes, as seen in recent MHD simulations \citep{Tchekhovskoy11,McKinney12}, corresponds to tens/hundreds of the gravitational radius light-crossing times, $\sim 100 \, r_g/c \sim 10^6 \, (M_{BH}/10^{9} M_{\odot})$\,s. This is of the correct order of magnitude assuming $M_{BH} \simeq 2 \times 10^{10} \, M_{\odot}$ suggested by \citet{Valtonen08} for OJ~287. Even so, in such a case the year-like periodicity claimed for radio and optical light-curves of other blazars (most likely hosting smaller black holes) would remain a puzzle, unless lower black hole spins are considered, allowing to accommodate longer timescales for smaller black hole masses.

\begin{acknowledgements}
We acknowledge the support by the Polish National Science Centre through the grants: DEC- 2012/04/A/ST9/00083 and 2013/09/B/ST9/00599.
We are grateful to C. Porowski,  T. Szymanski,  D. Jableka,  E. Kuligowska, J. Krzesinski, A. Baran  and A. Kuzmicz for carrying out some of observations. We gratefully acknowledge Marian Soida for his valuable assistance in the computational part of the work. We  thank the anonymous referee whose comments and suggestions greatly improved the analysis. This paper has made use of up-to-date SMARTS optical/near-infrared light curves that are available on line\footnote{ (\texttt{www.astro.yale.edu/smarts/glast/home.php})}

\end{acknowledgements}

\end{document}